# CONTROL OF MAGNETIC SUSCEPTIBILITY OF PROBIOTIC STRAIN *LACTOBACILLUS RHAMNOSUS* GG

*Svitlana Gorobets[1], Oksana Gorobets[2], Liubov Kuzminykh[3*]*

*Address:*
[1,2,3]National Technical University of Ukraine "Igor Sikorsky Kyiv Polytechnic Institute", Faculty of Biotechnology and Biotech Engineering, Department of Bioenergy, Bioinformatics and Environmental biotechnology, pr. Peremohy, 37, build. 4 and 18, postal code 03056 Kyiv, Ukraine, +380 (044) 204-97-79; +380 (044) 204-99-37
ORCID: Svitlana Gorobets 0000-0002-5328-2959; Oksana Gorobets 0000-0002-2911-6870

*Corresponding author: PhD-student Liubov Kuzminykh, e-mail: l.kuzminykh-2022@kpi.ua; ORCID: 0000-0002-6358-2883

**ABSTRACT**

The paper investigates the increase in the natural magnetically controlled properties of probiotic microorganisms *Lactobacillus rhamnosus* GG (LGG) and their ability to form magnetic sensitive inclusions (MsI). The magnetic susceptibility of LGG was increased by modifying the nutrient medium composition and by cultivating the probiotic culture in a constant magnetic field. Therefore, this study can be useful for their further use as magnetically controlled vectors, since it is being extensively researched for the use in targeted drug delivery in cancer treatment, for the prevention of chemotherapy complications, etc.
The results of this study indicate that growing microorganisms with natural magnetically-guided properties in a modified medium with iron chelate and in an external magnetic field leads to an increase in the magnetic susceptibility of LGG by 1,8 and 2,6 times, respectively, compared with the control. The best magnetic susceptibility was recorded for LGG suspensions, which were grown in a constant magnetic field and cultured in modified medium with iron chelate. The LGG suspension, grown both in a constant magnetic field and on a modified medium with iron chelate, had the highest magnetic susceptibility and was 4.8 times larger than the control.

## INTRODUCTION

Nowadays, it is hard to overestimate the role of probiotic microorganisms. Probiotics are used for treating and preventing the diseases of the gastrointestinal tract (**Parian et al., 2018**), strengthening the immune system (**Mendi et al., 2016**), overcoming the effects of antibiotic therapy (**Saarela et al., 2000**), for the prevention and treatment of tumor diseases (*Bedada et al., 2020; Lamichhane et al., 2020*), as vectors for targeted drug delivery (**Duong et al., 2019**). In particular, studies show a number of advantages while using probiotic microorganisms as vectors comparing to other microorganisms that can have toxic effects on the body (**Laliani et al., 2020; Zhou et al., 2018**). The use of probiotics as vectors for targeted drug delivery has a number of advantages, in particular, the non-invasiveness of probiotic cultures and the affinity for tumor tissue of some probiotic microorganisms (**Bedada et al., 2020**). Probiotic microorganisms can specifically target a tumor, causing intra tissue regression (**Tangney, 2010**). Microorganisms that synthesize biogenic magnetic nanoparticles (BMNs) or magnetosomes can be used as magnetically-guided vectors to transfer a large number of therapeutic substances, in particular antibiotics, antibodies, siRNAs, plasmid-based vaccines, anticancer agents - doxirubicin, cytarabine, etc. (**Kuzajewska et al., 2020**). Magnetotactic bacteria (MTB) are used with caution for such purposes because their effect on living organisms is not fully studied and because of the lack of clinical trials and regulatory documentation (**Mathuriya, 2015**). Also, MTB is difficult to cultivate and maintain their viability, since their habitat is significantly different from the internal environment of the human or animal organism (**Müller et al., 2020**). The genetic apparatus of BMNs biosynthesis is unique in all kingdoms of living organisms, as shown by genetic analysis (**O.Y. Gorobets et al., 2014**). For many years BMNs has been experimentally studied mainly in connection with the ideas of magnetotaxis and magnetoreception. BMNs of non-magnetotactic microorganisms has been studied in comparison with BMNs of MTB (**S. V. Gorobets 2012**). Whereas BMNs of non-magnetotactic microorganisms are also a source of their own strong gradient magnetic fields (**Gorobets and Koralewski 2017**). A number of probiotic and non-magnetotactic microorganisms can produce BMNs (**Horobets et al., 2014**). As a result, the use of probiotics that produce BMNs and magnetosensitive inclusions (MsI) is a promising solution for this class of problems (**Vainshtein et al., 2002, 2014**).
The strain of lactic acid bacteria *Lactobacillus rhamnosus* GG or *L. rhamnosus* ATCC 53103 (LGG) is a part of a number of probiotic drugs, including those that can be given to infants (**Buyukeren et al., 2020**). It is one of the most studied strains of probiotics due to its valuable properties - the microorganism colonizes the gastrointestinal tract and thus protects against pathogens, increases the body's resistance to infections (**Capurso, 2019**). LGG is used to prevent pathologies and reduce complications after tumor chemotherapy (**Banna et al., 2017**), thus increasing the effectiveness of treatment. LGG protects against pathogenic microorganisms that cause stomach disorders and diarrhea (**Petrova et al., 2016; Tytgat et al., 2016**), including *Clostridium difficile* is a pathogen that causes nosocomial infection (**Johnson et al., 2012**). LGG reduces the expression of cytokines and thus reduces inflammation and inhibits tumor growth in colon cancer (Caco-2) (**Li et al., 2020; Orlando et al., 2016**). It is known that LGG is used as a vector in targeted drug delivery in the treatment of pathologies and tumors (**Bedada et al., 2020; Tangney, 2010**). LGG produces biogenic silver nanoparticles (**Mousavi et al., 2020**), which are used in targeted drug delivery (**Radaic et al., 2020**). Bacteria from the genus *Lactobacillus* synthesize MsI when grown on special media with the addition of ferric ions (**Ariskina, 2003**).
Also, magnetically-guided vectors based on LGG can have additional advantages, as they can control their velocity to the tumor site, and they are well concentrated in the desired area (**Mokriani et al., 2021**). It is known that tumors can produce an increased amount of BMNs compared to healthy tissues (**Brem et al., 2006; Chekchun et al., 2011; S. V. Gorobets, 2017**). Magnetically-guided vectors are more efficiently attached to tumor cells and accumulate in the target area due to magnetic dipole – dipole interactions with tumors, which creates the required therapeutic effect (**Y. I. Gorobets et al., 2022; Mikeshyna et al., 2018**). Magnetically-guided vectors in the tumor area can be a magnetic material for therapeutic magnetic hyperthermia, which together with the therapeutic effect of the delivered drugs, can increase the effectiveness of treatment several times (**Felfoul et al., 2016; S. V. Gorobets et al., 2013**). Thus, the study of ways to control the magnetic susceptibility of LGG is relevant for the creation of magnetically-guided vectors based on the natural magnetically-guided properties of LGG.

**MATERIALS AND METHODS**

**Cultivation of microorganisms**

The probiotic lyophilized drug Acidolac® (1 sachet of the drug contains 4x10⁹ CFU *Lactobacillus rhamnosus* GG) was used for the study. Pure cultures of microorganisms were isolated and cultured on MPC-2 medium. The microorganisms were cultured on the standard agar medium and modified medium with the addition of iron chelate for 48 hours at a temperature of 36 ± 1 ºC. LGG was grown on the MRS medium prepared according to a standard recipe. The modified medium was prepared to enhance the natural magnetically-guided properties with the addition of iron chelate at a concentration of 64 mg / L. The media were prepared according to the recipe and sterilized by pressure in a steam sterilizer at 121 ºC for 15-20 minutes. The control group of microorganisms (LGG contr.) was grown on a standard medium under the standard conditions. LGG was grown to enhance the natural magnetically-guided properties with modification of the cultivation conditions: grown on a standard medium in a constant magnetic field (LGG + M), grown on a modified medium with the addition of iron chelate (LGG + Fe), grown on a modified medium with the addition of iron chelate in a constant magnetic field (LGG + Fe + M). The purity of the culture LGG was checked using by microscopic examination.

**Research of microorganisms with a two-magnet system**

The magnetic susceptibility of the suspension of LGG was investigated using a system of two magnets according to the method described (**Wosik et al., 2018**). A drop of suspension with a culture of microorganisms is applied to the glass located on the contact surface of the system of two permanent magnets. The system of two permanent magnets is created so that the highest gradient of the magnetic field is created in the area of their joint. The magnetic particles in the suspension move to the contact line of the magnets, forming a strip. Magnetic susceptibility was studied in two ways: concentrating the bacterial suspension on the contact line to determine the width of the formed band and determining the velocity of cell agglomerates to the contact line of the system of two permanent magnets.

**Bandwidth measurements of an LGG cell suspension formed at the contact surface of a system of 2 permanent magnets**

An aqueous suspension of probiotic strain LGG was applied to a 0.13 mm thick cover slide, which was placed on a two-magnet system until the liquid was completely dry to measure the width of the strip formed. After drying the product, the bandwidth of the suspension was evaluated using a digital microscope and a microline. The series of digital microscope images were taken and the average width of the formed strip was determined using the IrfanView software.

**Determination of the velocity and the average magnetic susceptibility of particles of the LGG suspension**

A drop of LGG suspension was applied to a glass placed on the contact surface of the system of two permanent magnets. The movement of conglomerates of LGG suspension was recorded on a video using a digital microscope. The video was filmed in several repetitions for the reliability of the results. Microorganisms were taken from different parts of Petri dish, a suspension was prepared and individual videos were made moving to the line of contact of the system of two permanent magnets. The experiment was repeated several times with separate generations of LGG cultures to obtain reliable results. An average of 15 videos were made for each type of microorganism cultivation. Particles with similar sizes were selected, the motion of which was clearly visible in the video, for the experiments. The number of particles examined was: LGG contr. - 57 particles, LGG + M - 78 particles, LGG + Fe - 83 particles, LGG + Fe + M - 139 particles. Particles were selected that were clearly visible, 40-110 µm in size, the average particle size was approx. 70 µm. The magnetic susceptibility of the LGG suspension was determined using Python and the formulas in the section of Results and Discussion.

Program IrfanView has been determine the particle movement (µm) to the contact line of the magnets and the particle diameter (µm). Time (s) of movement of the particle was determined by dividing the video into frames by Free Video to JPG Converter software. The average velocity (µm/s) of the particles was determined by dividing the path (µm) by the time (s) of movement of the particles to the contact line of the magnets. Table 1 presents the physical parameters were taken to determine the magnetic susceptibility of the LGG suspensions.

**Table 1** Physical parameters of the system, which were used to calculate the average magnetic susceptibility of suspensions in the study with a system of two magnets.

| Parameter | Dimension | Value, units |
|---|---|---|
| Saturation magnetization of magnets in a system of two permanent magnets | $M_s$ | 1600 G [a] |
| Dynamic viscosity of water | $\eta$ | 0.01 g/(cm·s) |
| Magnet length | $a$ | 0.65, cm |
| Glass thickness | $h$ | 0.13, mm |

Note: a - the unit is part of the Gaussian system of units or CGS-EMU, 1 G = $10^{-4}$ tesla (SI system).

**RESULTS AND DISCUSSION**

**Determination of bandwidth and velocity of suspension particles LGG**

The microscopy of the studied of probiotic strain LGG is shown in Figure 1.

**A**   **B**

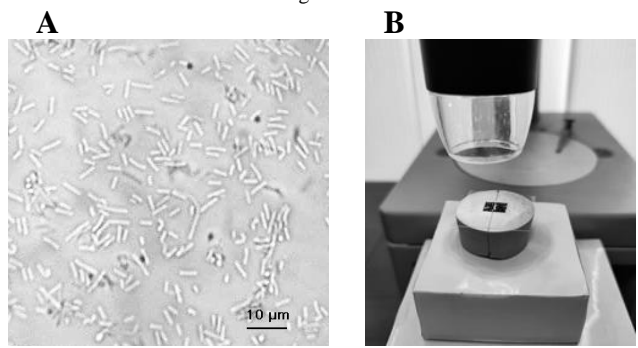

**Figure 1** Probiotic bacteria *L. rhamnosus* GG, cultured from the Acidolac on MRS medium (A). Digital microscope examination of the magnetophoretic mobility of *L. rhamnosus* GG: a system of two permanent magnets is presented, where a cover glass with a culture concentrating on the contact line is located on top (B).

Fig. 2 shows a preparation from a suspension of LGG, placed over a system with 2 magnets. The images show the conglomerates of bacterial cells are located along the contact line of the magnetic system.

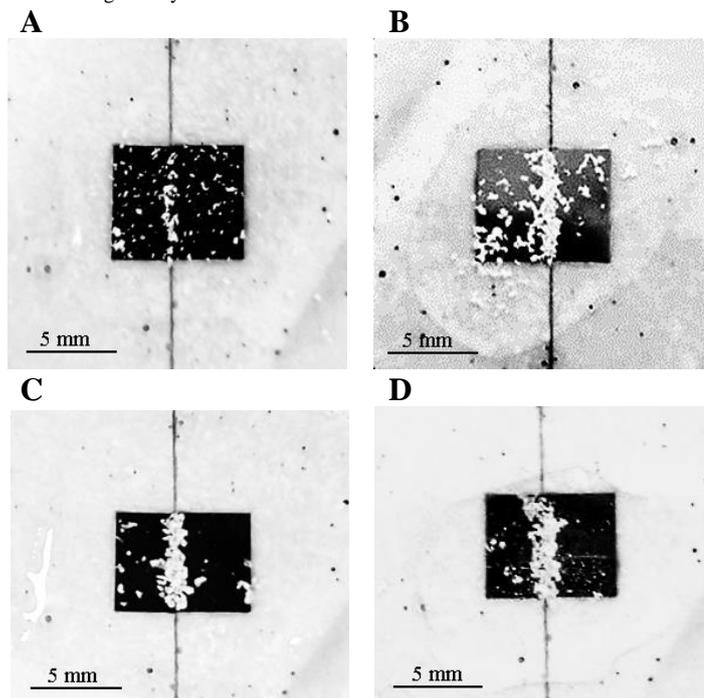

**Figure 2** Stripes from a suspension of probiotic strain *L. rhamnosus* GG, which formed above the contact line of the system of two magnets: LGG contr. (A), LGG +M (B), LGG + Fe (C), LGG + Fe + M (D).

Wider and less branched strip form above the contact line when the natural magnetically-guided properties of the LGG culture are enhanced by iron ions and a constant magnetic field. The strip that is formed after cultivation under the standard conditions (LGG contr.) is thin, indistinct, with gaps and branches, many particles are not on the line of contact and do not move towards the line. The strip is formed by cultivation in a constant magnetic field (LGG + M) is distinct along its entire length, but rather branched; a certain number of suspension particles do not move towards the contact line. The strip that is formed after the cultivation on a modified medium with iron chelate (LGG + Fe) is somewhat narrower comparing to the strip formed by LGG + M. The strip formed for LGG + Fe on the contact line of two magnets is distinct, rather uniform, the particles are concentrated on the line, therefore, they form a denser strip with few branches; there are few particles that remain outside the strip and do not move towards the contact line. The strip for LGG + Fe is narrower than the strip for LGG + M due to the fact that the particles are more densely concentrated on the contact line. The strip formed during cultivation on a modified medium with the addition of iron chelate and in a constant magnetic field (LGG + Fe + M) is the widest strip compared to the strips formed for other cultivation conditions. A distinct wide strip is formed for LGG + Fe + M rather homogeneous, has few branches; few particles remain outside the strip and do not move towards the contact line.

Figure 3 shows the results of the average velocity of particles to the line of contact of the system of two magnets. The average velocity (μm/s) of the particles was obtained by dividing the path of the particle by time.

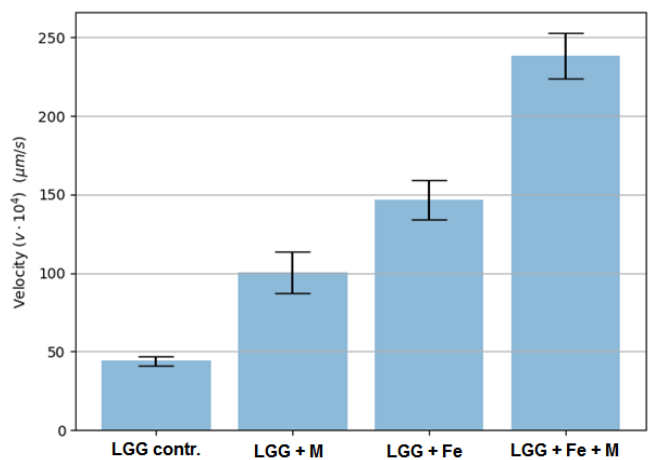

**Figure 3** Dependence of the average velocity of the particles of the suspension of *L. rhamnosus* GG depending on the cultivation conditions. The error was determined using the standard deviation from the mean (SDM).

The results in this paper show that the speed of the suspension particles in the two-magnet system increased as the growing conditions changed and the culture medium was modified.

**Calculation of the magnetic susceptibility of the suspension LGG**

The main idea of modifying the magnetic properties of bacteria with natural magnetically-guided properties is to grow these bacteria under special conditions, namely with the addition of iron chelates to the environment and under the application of an external constant magnetic field during cultivation. This idea is based on experimental data on the influence of cultivation conditions on the amount of BMNs in magnetotactic bacteria. The BMNs of magnetotactic bacteria increases up to several times due to an increase in the concentration of iron ions in the cultivation medium and the application of an external magnetic field with an induction of about 0.1 T during cultivation. The magnetic susceptibility of biomass of magnetotactic bacteria also increases (**Wang *et al.*, 2009**). According to **O. Y. Gorobets *et al.* (2014); Mokriani *et al.* (2021),** it can be assumed that the same influencing factors that regulate the amount of BMNs in magnetotactic bacteria will similarly affect the number of BMNs and the magnetic susceptibility of non-magnetotactic bacteria with natural magnetically-guided properties. In addition, the novelty of this work is the choice of non-magnetotactic bacteria with natural magnetically-guided properties that are used as vectors for drug delivery (**Bedada *et al.*, 2020; Zhou *et al.*, 2018**).

In recent years, methods for creating strong magnetic fields have been intensively developed using systems of highly anisotropic permanent magnets connected in such a way that the directions of the magnetization vectors of these magnets are different at the boundary between them. In this case, strong magnetostatic fields are created at the "sharp angles" of each of the magnets, having a logarithmic divergence at certain points in space.

This divergence is not distorted due to the bending of the magnetization vector, if the material from which the magnet is made of has a large magnetic anisotropy. **Akhiezer *et al.* (1968) *and* Samofalov *et al.* (2004, 2013)** calculated the divergence of the tangential component of the magnetic field outside a uniformly magnetized rectangular parallelepiped. The magnetostatic field increases logarithmically when approaching the edges of the parallelepiped and can exceed the value $4\pi M_s$ that is the limiting value for permanent magnets of traditional designs.

Fig. 4 presents an option of implementation of this idea. The combination of several highly anisotropic ferromagnets of various shapes with different directions of magnetization in one magnet to create strong magnetic fields near their common "sharp angle".

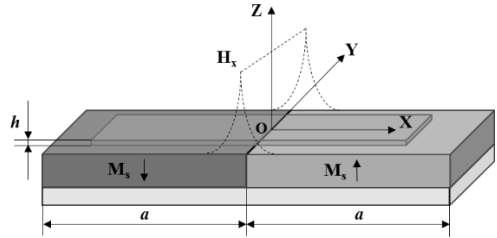

**Figure 4** Scheme of Kittel's open domain structure of two magnets.

Fig. 4 presents a magnet where two ferromagnets are connected in such a way that one of them is magnetized along the axis $OZ$ and the other one in the opposite way. The expression for the tangential component of the magnetostatic field for the Kittel's domain structure of two magnets has the form (Akhiezer et al., 1968; Samofalov et al., 2004, 2013):

$$H_x^{(m)}(x,y,z) = M_s \left[ \ln(x^2 + 2ax + a^2 + z^2) + \ln(x^2 - 2ax + a^2 + z^2) - 2\ln(x^2 + z^2) \right] \quad (1)$$

Expression (1) contains a logarithmic divergence at $z = 0$ and $x \to 0$.

Therefore, a magnet of this shape, made of highly anisotropic material, can be used to create strong magnetic fields in localized volumes of space. The normal component of the magnetostatic field for the Kittel's domain structure of two magnets has the form (Samofalov et al., 2004, 2013):

$$H_z^{(m)}(x,y,z) = 2M \left( \operatorname{arctg}\left(\frac{a+x}{z}\right) - \operatorname{arctg}\left(\frac{a-x}{z}\right) - \operatorname{arctg}\left(\frac{x}{z}\right) \right) \quad (2)$$

Analysis of the existing theories of particle deposition in a gradient magnetic field has revealed that the theory of magnetophoresis of para- and diamagnetic particles in high-gradient fields has been quite fundamentally developed. The choice of a model to describe the efficiency of the magnetic separator depends on the size of the particles that must be isolated from the working medium. There are two approaches which depend on particle size (more or less some critical size), the particle approach and the continuous medium approach. The continuous medium approach is used to model the separation of particles, usually less than one micron in size. The statistical method is used when it is impossible to estimate the spontaneous Brownian motion of each particle, and the suspension is considered as a continuous medium. The particle approach gives good results for particles of sufficient size to be neglected Brownian motion. The model using the particle approach was first developed in (**Friedlander *et al.*, 1981; S. V. Gorobets & Mikhailenko, 2014; Plyavin' & Blum, 1983; Vincent-Viry *et al.*, 2000**). This model takes into account only the gradient magnetic force $\vec{F}_m$ and the force of hydrodynamic resistance $\vec{F}_{st}$. The force of inertia is neglected, and the balance of forces is written as in the following form:

$$\vec{F}_m = \vec{F}_{st} \quad (3)$$

$$\vec{F}_m = \frac{1}{2}\chi \operatorname{grad}\vec{H}^2 V_p \quad (4)$$

$\vec{H}$ – the strength of the magnetic in the working volume of the magnetic separator, $\chi$ – effective magnetic susceptibility, which is equal to the difference between the susceptibilities of the liquid and the particle $\chi = \chi_p - \chi_f$, $V_p$ - the volume of the particle.

The gradient magnetic force is directly proportional to the gradient of the magnetic energy density. Thus, if the field is locally homogeneous, it means that no force acts on the particles. A non-zero gradient magnetic force exists in an inhomogeneous magnetic field. If the motion can be characterized by small Reynolds numbers, the Stokes force acting on a spherical particle has the form:

$$\vec{F}_{st} = 6\pi \eta b \vec{v} \quad (5)$$

Assuming that the shape of a particle trapped in an inhomogeneous magnetic field is a sphere of $b$ radius which is moving in a liquid with dynamic viscosity $\eta$, $\vec{v}$ the velocity difference between liquid and particle in a stationary medium.

The capture process is determined as a result of the competition between the magnetic force $\vec{F}_m$ and the Stokes force $\vec{F}_{st}$ (**Friedlander *et al.*, 1981**).

The magnetic field strength is described by expressions (1), (2) in the case when an inhomogeneous magnetic field is created by a system of two magnets shown in Fig. 4. We assume that a thin layer of liquid on the surface of the system of magnets is stationary, then the velocity of spherical particles in the liquid on the surface of the system of magnets along the axis, and the characteristic radius of the particle. Equation (3) can be converted to the next dimensionless form:

$$\frac{dX}{d\tau} = f(X,Z) \quad (6)$$

where dimensionless coordinates $X = \frac{x}{a}$, $Y = \frac{y}{a}$, $Z = \frac{z}{a}$ and $\tau = \frac{\chi M_s^2 b^2}{9a^2\eta} \cdot t$

and notation are introduced:

$$f(X,Z) = \frac{d}{dX}\left\{4\left(\arctan\left(\frac{1+X}{Z}\right) - \arctan\left(\frac{1-X}{Z}\right) - \arctan\left(\frac{X}{Z}\right)\right)^2 + \ln\frac{(1+2X+X^2+Z^2)(1-2X+X^2+Z^2)}{(X^2+Z^2)^2}\right\} \quad (7)$$

As a result of integrating the equation (7) we find the dimensionless time $\tau_X$ required for the particle to move in the magnetic field of the magnet system from a point with dimensionless coordinate $-X$ to the line of contact of two magnets with $\frac{b}{a}$ dimensionless coordinate, as a function of dimensionless particle coordinates $Z$:

$$\tau_x\left(X, Z, \frac{b}{a}\right) = \int_{-X}^{\frac{b}{a}} \frac{d\tilde{X}}{f(\tilde{X}, Z)} \quad (8)$$

The suspension of particles sedimented in the gravitational field before being introduced into the magnetic field, therefore, $Z = \frac{b+h}{a}$ the dimensionless coordinate of the particle was chosen for the calculation, where $h$ is the thickness of the glass on the surface of the system of two magnets. This means that the center of the spherical particle is at a height above the surface of the magnet system equal to the radius of the particle with the addition of the thickness of the glass. It can be assumed with great accuracy that $Z = 0$ if $b + h \ll a$.

$$\tau_x\left(X, \frac{b}{a}\right) = \int_{-X}^{\frac{b}{a}} \frac{d\tilde{X}}{f(\tilde{X}, 0)} \quad (9)$$

Calculation $\left\langle \frac{dX}{d\tau} \right\rangle$ of the average dimensionless velocity of the particle along $0x$ the axis in $\left[-X; \frac{b}{a}\right]$ the region based on expression (8):

$$\left\langle \frac{dX\left(X, Z, \frac{b}{a}\right)}{d\tau} \right\rangle = \frac{X}{\tau_x\left(X, Z, \frac{b}{a}\right)} \quad (10)$$

The dependence of $V = \left\langle \frac{dX\left(X, Z, \frac{b}{a}\right)}{d\tau} \right\rangle$ average dimensionless velocity on $X$ when $Z=0$ is presented in Fig. 4.

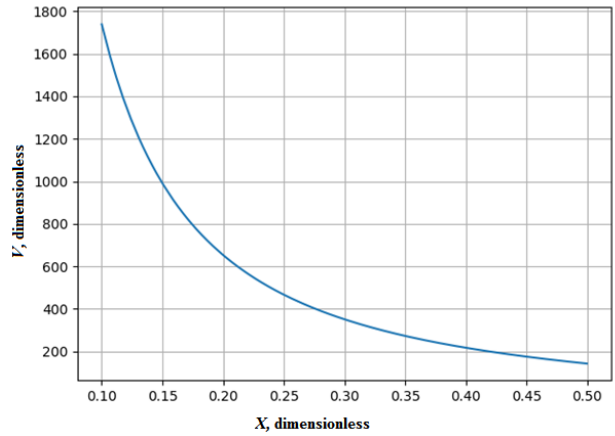

**Figure 5** Dependence of the average dimensionless velocity of the particle on $X$ at $Z=0.001$ and at $\frac{b}{a} = 0.001$ calculated by (10).

The average velocity of a particle measured in seconds, is determined through the dimensionless average velocity by the formula:

$$\left\langle \frac{dx}{dt} \right\rangle = \frac{\chi M_s^2 b^2}{9a\eta} \left\langle \frac{dX\left(X, Z, \frac{b}{a}\right)}{d\tau} \right\rangle \quad (11)$$

The effective magnetic susceptibility of a particle moving in a liquid on glass on the surface of a system of two magnets can be determined from the last expression:

$$\chi = \frac{9a\eta}{M_s^2 b^2} \cdot \frac{\left\langle \frac{dx}{dt} \right\rangle}{\left\langle \frac{dX\left(X, Z, \frac{b}{a}\right)}{d\tau} \right\rangle} \quad (12)$$

The dimensionless coordinate of the particle is chosen for calculation, as already mentioned: $Z = \frac{b+h}{a}$. The results of the calculation of the average effective magnetic susceptibility of particles depending on the cultivation conditions by formula (12) are presented in Fig. 6.

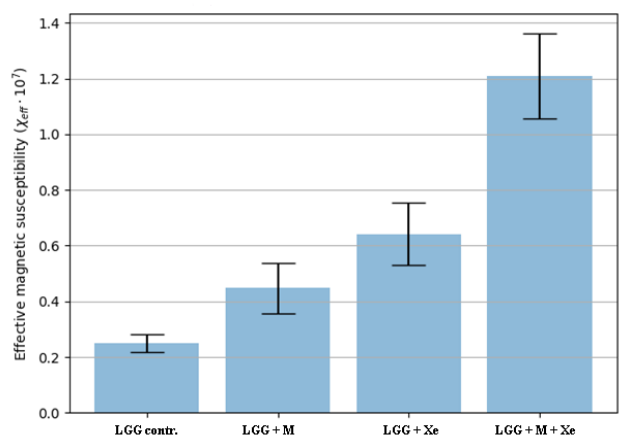

**Figure 6** The average effective magnetic susceptibility of the particles of the *L. rhamnosus* GG suspension depending on the cultivation conditions according to formula (12). The error was determined using the SDM.

**CONCLUSION**

The results show that the speed of the suspension particles in the two-magnet system increased as the growing conditions changed and the culture medium was modified. The effective magnetic susceptibility for the LGG contr. sample is $(0.25 \pm 0.03) \cdot 10^{-7}$. The magnetic susceptibility of the LGG bacterial culture when grown in a constant magnetic field is $80\% \pm 6\%$ and, accordingly, 1.8 times larger; when modifying the environment with iron chelate by $156\% \pm 10\%$ and, accordingly, 2.6 times larger, and cultivation in a constant magnetic field by $384\% \pm 6\%$ and, accordingly, 4.8 times larger than the magnetic susceptibility for the control sample. Thus, modification of the growth medium with iron chelate and/or cultivation in a constant magnetic field improves the ferrimagnetic properties of the bacterial culture LGG, which can be useful in creating vectors of magnetically-guided delivery. A significant increase in the magnetic susceptibility of LGG bacterial culture is observed when both methods are used together: modification of the medium with iron chelate and cultivation in a constant magnetic field.